\documentclass[10pt,conference]{IEEEtran}

\usepackage{xspace}
\usepackage[colorlinks=true,urlcolor=black,linkcolor=black,citecolor=black]{hyperref}
\usepackage{booktabs}
\usepackage{xspace}
\usepackage{balance}
\usepackage{microtype}
\usepackage{listings,lst-kconfig,multicol}
\usepackage{xcolor}
\usepackage{url}
\usepackage{relsize}
\usepackage{csquotes}
\usepackage{threeparttable}
\usepackage{tabularx}
\usepackage{algorithm,algorithmicx,algpseudocode}
\usepackage{cite}
\usepackage[normalem]{ulem}

\newcommand{\longv}[1]{}
\newcommand{\algo}[1]{}

\newcommand{\secref}[1]{Sec.\,\ref{#1}}

\newcommand{\figref}[1]{Fig.\,\ref{#1}}

\newcommand{\tabref}[1]{Table\,\ref{#1}}
\newcommand{\Figref}[1]{Figure\,\ref{#1}}

\newcommand\parhead[1]{\vspace{0.5mm}\noindent\textbf{{#1}}.}

\def\inline{\lstinline[basicstyle=\ttfamily\fontsize{8}{8}\selectfont,,keywordstyle={}]}

\newcommand{\hashif}{\inline{\#if}\xspace}
\newcommand{\xconfig}{\textit{xconfig}\xspace}
\newcommand{\Xconfig}{\textit{Xconfig}\xspace}
\newcommand{\ecos}{\textsc{eCos}\xspace}
\newcommand{\toolname}{\textsc{ConfigFix}\xspace}
\newcommand*\circlenum[1]{\tikz[baseline=(char.base)]{
    \node[shape=circle,draw,inner sep=2pt, fill=gray!20] (char) {#1};}}

\usepackage{tikz}
\usepackage{calc}
\newlength{\leftLength}
\newcommand{\summary}[2]{
\setlength{\leftLength}{0.5cm + (\widthof{{\small \textbf{#1}}}/2)}
\vspace{+.1cm}\noindent\begin{tikzpicture}
\node[align=center,draw,thin,minimum width=\columnwidth,inner sep=2.2mm] (titlebox)%
{\parbox{0.95\columnwidth}{\vspace*{0.25ex}\noindent\emph{#2}}};\node[fill=white] (W) at ([xshift=\the\leftLength] titlebox.north west) {{\small \textbf{#1}}};%
\end{tikzpicture}\vspace{+.1cm}}

\title{ConfigFix: Interactive Configuration\\Conflict Resolution for the Linux Kernel\\[+.5cm]{\large Experiences from a Decade of Reverse-Engineering the Semantics from\\[-.5cm]the Linux Kernel Configurator and Realizing Intelligent Configuration Support}\vspace{-.2cm}
}

\author{
	\IEEEauthorblockN{Patrick Franz\IEEEauthorrefmark{1}, Thorsten Berger\IEEEauthorrefmark{2}\IEEEauthorrefmark{1}, Ibrahim Fayaz\IEEEauthorrefmark{3}, Sarah Nadi\IEEEauthorrefmark{4}, Evgeny Groshev\IEEEauthorrefmark{1}}
	\IEEEauthorblockA{\IEEEauthorrefmark{1}Chalmers\,$|$\,University of Gothenburg \IEEEauthorrefmark{2}Ruhr University Bochum \IEEEauthorrefmark{3}VecScan AB (Vector Sweden) \IEEEauthorrefmark{4}University of Alberta}
}

\begin{document}

\maketitle

\begin{abstract}
\looseness=-1
Highly configurable systems are highly complex systems, with the Linux kernel arguably being one of the most well-known ones. Since 2007, it has been a frequent target of the research community, conducting empirical studies and building dedicated methods and tools %
for analyzing, configuring, testing, optimizing, and maintaining the kernel in the light of its vast configuration space.
However, despite a large body of work, mainly bug fixes that were the result of such research made it back into the kernel's source tree. Unfortunately, Linux users still struggle with kernel configuration and resolving configuration conflicts, since the kernel largely lacks automated support.
Additionally, there are technical and community requirements for supporting automated conflict resolution in the kernel, such as, for example, using a pure C-based solution that uses only compatible third-party libraries (if any).

\looseness=-1
With the aim of contributing back to the Linux community, we present \toolname, a tooling that we integrated with the kernel configurator, that is purely implemented in C, and that is finally a working solution able to produce fixes for configuration conflicts. In this experience report, we describe our experiences ranging over a decade of building upon the large body of work from research on the Linux kernel configuration mechanisms as well as how we designed and realized \toolname while adhering to the Linux kernel's community requirements and standards. %
While \toolname helps Linux kernel users obtaining their desired configuration, the sound semantic abstraction we implement provides the basis for many of the above techniques supporting kernel configuration, helping researchers and kernel developers.

\end{abstract}


\section{Introduction}
\label{sec:introduction}
\looseness=-1
\noindent
The Linux kernel is the world's largest software development pro\-ject\,\cite{linux_largest} by the number of its contributors. Being highly versatile, the kernel operates in a diversity of environments, ranging from Android phones to large supercomputer clusters. As such, it is not only a highly successful operating-system kernel, but also a \textit{highly configurable system}\,\cite{sincero:08:linux}---nowadays boasting 28 million lines of code\,\cite{linux_stats} and over 15,000 configuration options (a.k.a., \textit{features}\,\cite{passos.ea:2018:tse,berger.ea:2015:feature}). To this end, the kernel relies on mechanisms known from the fields of software product lines\,\cite{apel.ea:2013:fospl,berger.ea:2020:emse}, model-driven engineering\,\cite{brambilla2017model}, and software configuration\,\cite{berger2014variability}. Specifically, the Linux kernel includes a configurable build system\,\cite{berger.ea:2010:featuretocode}, preprocessor-enabled variation points, a model-based representation of configuration options and their constraints (a.k.a., \textit{variability model})\,\cite{berger2013study,nadi.ea:2015:tse}, and an interactive configurator tool\,\cite{sincero:08:linux}.
Being completely open source, with a vast evolution history available, researchers have studied many different aspects of it, including software evolution\,\cite{israeli2010linux,tu2000evolution,antoniol2002analyzing,passos.ea:2018:tse} and software maintenance\,\cite{TianLinuxBugFix12,israeli2009characterizing,abal.ea:2014:varbugs,jiang2013will} aspects, as well as its configuration mechanisms---the focus of this paper.

\looseness=-1
Studies of the Linux kernel's configuration mechanisms started back in 2007\,\cite{sincero.ea:osspl,sincero:08:linux,tartler.ea:2009:fosd}, followed in 2010 by our and other researchers' studies of its variability modeling language Kconfig and its variability model\,\cite{she2010variability,berger2010variability,berger2013study}. Examples include the evolution of this model\,\cite{lotufo:splc:2010}, the co-evolution and consistency of variation points\,\cite{passos.ea:2016:coevol,passos.ea:2018:tse,NadiWCRE2011,NadiCSMR2012,
kastner2011variabilityaware,TartlerUndertaker2011}%
, as well as the synthesis of variability models from code\,\cite{she2011reverse,nadi.ea:2014:constraints,nadi.ea:2015:tse}. 
Despite all the above research efforts related to configuration, users of the kernel did not benefit directly yet. They still struggle with creating their desired configuration\,\cite{hubaux_study}, given the huge configuration space and intricate constraints among features.
Beyond a simple and very limited support for choice propagation%
, the configurator does not offer any intelligent support for resolving configuration conflicts---for instance, enabling a feature requires transitively changing many other features. As such, achieving the desired configuration can be laborious and error-prone, which is unfortunate given all the work in the research community, which never made it back into the kernel. 

\begin{figure}[t]
\centering
	\includegraphics[width=\columnwidth]{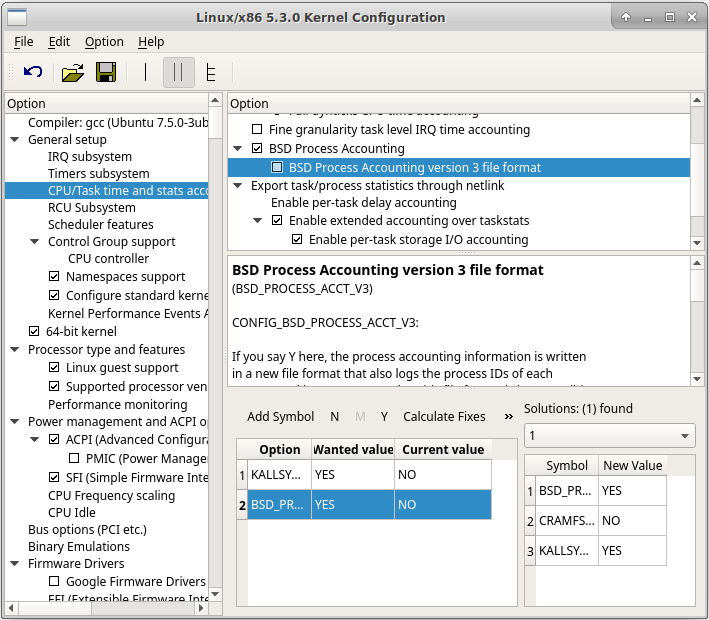}
	\vspace{-.5cm}
	\caption{The kernel configurator extended with \toolname} %
	\label{fig:xconfig}
	\vspace{-.5cm}
\end{figure}

To the best of our knowledge, only few contributions originating from research made it back into the kernel, and mainly in the form of bug fixes\,\cite{LawallDriver2008,TartlerUndertaker2011,NadiWCRE2011}. %
While there are tools such as Coccinelle\,\cite{Lawal10YrsUSENIX18} and Undertaker\,\cite{UndertakerUsenix2014} that become known and often individually adopted among kernel developers, none were formally integrated into the kernel codebase or are listed as a necessary kernel tool.
In 2015, the situation was about to change with the Kconfig-sat initiative\,\cite{kconfig-sat-wiki}, where kernel developers recognized the need and got in touch with researchers working on kernel configuration studies, including us. Given our experience with studying the Linux kernel's configuration information and developing tools to analyze it, we decided it was time to give back to the Linux community and try to practically integrate these techniques into the official Linux kernel configurator.
In fact, implementing a sound translation of the variability model to propositional logics (which is a semantic abstraction) given the expressiveness and intricate semantics of the the Kconfig language (explained shortly), has long been an open problem, with multiple translations proposed\,\cite{she2010formal,undertaker,kconfigreader}, each with its own shortcomings\,\cite{elsharkawy.ea:2015:survey}.

\looseness=-1
We describe experiences covering more than a decade of different efforts on reverse-engineering the formal semantics from the Linux kernel configurator and implementing sound semantic abstractions. These are the prerequisite for many techniques and use cases supporting software configuration. After providing an overview on the Linux kernel's configuration tooling, research efforts on it, and their historical perspective, we introduce our tool \toolname, which offers intelligent configuration support for the Linux kernel configurator and realizes an evaluated, pure-C implementation of a semantic abstraction and a technique to resolve configuration conflicts. \toolname takes a current configuration (i.e., a selection of features with concrete values), a set of configuration constraints declared in the kernel's variability model, and a set of configuration options whose values the user wants to change, in order to calculate a set of \emph{fixes} to reach the desired configuration.
We built on our and others' prior work in the field to realize a translation of the kernel's variability model into propositional logics, to implement a configuration-conflict resolution algorithm relying on SAT solving, and to integrate both into the graphical configurator tool \xconfig.
Our work led to finally obtaining a viable solution integratable in the kernel's source tree. \toolname is freely available\,\cite{configfix}, together with details about its evaluation.

\section{Software Configuration and the Linux Kernel}
\label{sec:background}
\noindent
\looseness=-1
We briefly introduce the field of software configuration, the Linux kernel's configuration facilities, and the notion of configuration conflict.

\subsection{Software Configuration}
\noindent
\textit{Software configuration} is concerned with methods and tools to configure software, originally stemming from the field of product configuration, a subfield of AI%
\,\cite{hubaux.ea:2012:unifying,aimodernapproach}. The challenge is to obtain a configuration that meets end-user requirements, considering all constraints among the configuration options (i.e., \textit{features}). The configuration process is typically supported with an interactive configurator tool, offering support for propagating choices and resolving configuration conflicts. Software configurators\,\cite{bashroush.ea:2017:plcasetools,benavides2010automated} have been studied in many domains\,\cite{berger.ea:2020:emse,linden.ea:2007:practices,berger2013survey}, including configurable systems software (e.g., Linux kernel, %
eCos embedded operating system or 3D printer firmware\,\cite{sincero.ea:osspl,berger2013study,krueger2018marlin,berger.ea:2010:cdlnote,she2010formal}), automotive\,\cite{flores.ea:2012:GM}, avionics\,\cite{sharp1998reducing,lindohf.ea:2020:emse}, and telecommunication systems\,\cite{svahnberg.ea:1999:evolution}, embedded and safety-critical software\,\cite{toft.ea:2000:hpowen,krueger.ea:2008:homeaway,berger.ea:2014:industry,schmid.ea:2005:testo}, as well as web-based configuration\,\cite{abbasi.ea:2012:webconfigurator}.

\subsection{The Linux Kernel and its Configurator}
\noindent
\looseness=-1
The Linux kernel's
configurability aims at customizing the kernel beyond its core functionality of CPU \& memory management towards many different hardware architectures (ranging from embedded devices to supercomputer architectures) and including optional functionality, such as device or filesystem drivers.
Currently, over 15,000 configuration options (henceforth called \emph{features}\,\cite{passos.ea:2018:tse,berger.ea:2015:feature}), which come with intricate constraints among them, control
variation points in C source files using conditional compilation directives (e.g., \hashif) of the C preprocessor, as well as they control the inclusion of individual files in the build process. In addition to this static mechanism%
, many features also control \textit{loadable kernel modules} (e.g., network or USB drivers) that can be loaded dynamically at runtime.

\looseness=-1
Users configure the kernel interactively via its configurator, which exists in three variants. \Figref{fig:xconfig} shows the graphical configurator \xconfig. The other two variants target shell users. 
All features come with default values, and users can then assign values to the individual features according to their types and constraints, establishing a configuration. %
The features, their organization in a hierarchy, and their constraints are declared in files using the Kconfig language, which
are input to the configurator. %

\looseness=-1
\Xconfig supports basic validity checking of configuration choices (to prevent some constraint violation) as well as simple imperative choice propagation. The latter, given the absence of an intelligent reasoner, needs to be encoded with a dedicated imperative mechanism in Kconfig, which is error-prone and, given its imperative nature, cannot be used to resolve configuration conflicts.
In contrast, 
various open-source (e.g., FeatureIDE\,\cite{meinicke2017mastering}, Dopler\,\cite{dhungana.ea:2011:dopler}, eCos' configtool\,\cite{berger.ea:2010:cdlnote}) and commercial configurators (e.g., pure::variants\,\cite{beuche2004variants}, Gears\,\cite{krueger2013systems}), come with a reasoner.

\subsection{The Kconfig Language}

\looseness=-1
\noindent
At the core of the Linux kernel configurator is the language Kconfig---a domain-specific language for variability modeling.
Originally created for the Linux kernel, it has since been adopted by at least ten other open-source projects, such as BusyBox\,\cite{berger2013study}. A core challenge in the community was obtaining a sound logical representation of the main semantics of Kconfig as a prerequisite to develop analysis and configuration techniques.
However, as we will illustrate, Kconfig is surprisingly expressive with exceptionally intricate semantics. %

\parhead{Language Concepts}\label{sec:kconfigconcepts}
\looseness=-1
\noindent
Kconfig comes with a textual syntax and concepts known from feature modeling (a popular kind of variability modeling language)\,\cite{kang.ea:1990:foda,czarnecki.ea:00:generative,Nesic:2019:PFM:3338906.3338974}: a \emph{hierarchy of features}, different \emph{feature types}, \emph{feature groups} (e.g., OR, XOR or MUTEX groups), and cross-tree constraints\,\cite{berger2013study}. A feature model describes the set of all possible configurations as its main semantics.
Since feature modeling languages are typically limited to Boolean features and propositional constraints, they can easily be converted into propositional logics by implementing their semantics.

\looseness=-1
Kconfig's syntax and semantics go well beyond feature modeling. For scaling the variability model and configuration process, Kconfig incorporates concepts such as visibility conditions (to conditionally show whole subtrees), modularization concepts, derived defaults / derived features, hierarchy manipulation), and an expressive constraint language including comparison, arithmetic, and String operators. Interestingly, features can also inherit constraints of their parents in non-transparent ways.
Furthermore, Kconfig has a domain-specific vocabulary (main keywords) that fosters comprehension among Linux developers.

\looseness=-1
These concepts substantially complicate Kconfig's syntax and semantics. In fact, many intricate semantic interactions between different language elements exist---most notably between seven (sic!) language constructs to express constraints (\texttt{prompt}, \texttt{default}, \texttt{depends on}, \texttt{select}, \texttt{imply}, \texttt{visible if}, and \texttt{range}). For instance, a default value becomes a constraint when the feature is not visible, as determined by other constraints.
For further details, we refer to Kconfig's official documentation\,\cite{zippel:09:Kconfig} and our prior work\,\cite{she2010formal,berger2013study}, which includes seven pages of reverse-engineered denotational semantics for Kconfig.
Finally, Kconfig is continuously extended with language constructs, such as recently with the statement \texttt{imply}\footnote{\url{https://gitlab.freedesktop.org/panfrost/linux/commit/237e3ad0f195d8fd34f1299e45f04793832a16fc}} as a special case of the \texttt{select} statement used for imperative choice propagation.

\summary{Imperative Choice Propagation}{%
A frequent issue driving the complexity of the Kconfig semantics is that the developers incorporated imperative choice propagation---mainly through the \texttt{select} statement, which interacts with other Kconfig elements in intricate ways. This issue frequently complicated our and others' efforts realizing a sound semantic abstraction. It would have been better to keep the language cleaner and separately implement choice propagation via a reasoner, as \toolname does, but now accounting for the imperative choice propagation.
}

\looseness=-1
Features can be of different types: \texttt{bool}, \texttt{tristate}, \texttt{string}, \texttt{hex}, and \texttt{int}. Tristate features are used to control the binding mode of features and can have three values: \textbf{y} (yes, compile feature into kernel), \textbf{n} (no, do not compile feature) or \textbf{m} (mod, compile as loadable kernel module). Kconfig offers this feature type together with three-state logics that follows Kleene's rules for three-state logics\,\cite{kleene:1938:threestate}. Intuitively, the value of a tristate feature is encoded as the number 0, 1 or 2. The logical operators are then defined over numbers: \&\& returns the minimum and $||$ the maximum of the two operands, and ! returns 2 minus the operand.

In our experience, the complexity of Kconfig is a result of the design of its configurator tooling. Instead of performing expensive language engineering\,\cite{combemale2016engineering,Brambilla2012,schauss2017sle,lammel2018software,erdweg2013languageworkbenches} and adopting a configurator that comes with more intelligent reasoning capabilities\,\cite{bashroush.ea:2017:plcasetools}, we learned that the community prefers transparent and easily scriptable solutions as opposed to heavy machinery, such as off-the-shelf reasoners that are difficult to understand. We learned this preference from the discussion on the kernel mailing list preceding the introduction of Kconfig and its tooling. An alternative candidate was a  a configurator tool and language with built-in conflict-resolution support, which the community explicitly decided against, expressing these reasons.

\summary{Kconfig Language and Configurator Design}{%
Kconfig is a popular language, but surprisingly expressive, coming with intricate syntax and semantics. We learned that the kernel community preferred the script-style \xconfig and Kconfig over more systematically engineered tooling%
, mainly to be able to fully control and evolve the tooling.
}

\parhead{Language Semantics and Abstractions}\label{sec:translations}
\noindent
\looseness=-1
Motivated by the prospect of interesting empirical insights and being able to evaluate configuration- and variability-related research prototypes, researchers started looking into Kconfig and its tooling back in 2007\,\cite{sincero.ea:osspl,sincero:08:linux,tartler.ea:2009:fosd}. This was followed up with holistic studies of Kconfig, its tooling, and its models in 2010\,\cite{she2010variability,berger2010variability,berger2013study}. For instance, we were the first to formally describe the semantics of Kconfig, which allowed its translation into propositional logic both by ourselves and other researchers\,\cite{she2010formal}, which resulted in the first translation tool called LVAT\,\cite{lvat} (Linux Variability Analysis Tools), a tool suite written in Scala for analyzing Kconfig models.
Around the same time, Zengler and K{\"u}chlin also provided a translation of Kconfig into propositional logics\,\cite{zengler2010encoding,elsharkawy.ea:2015:survey}. %
Furthermore, the tool Undertaker\,\cite{undertaker} analyzes \texttt{\#ifdef} code in the Linux kernel to identify dead code blocks---variation points whose constraints conflict with the variability model. To this end, Undertaker came with a translation into propositional logics.
As part of the TypeChef infrastructure\,\cite{kastner2011variabilityaware}, K\"{a}stner implemented Kconfigreader\,\cite{kconfigreader} in Scala. %
Most recently, Fernandez-Amoros et al. also proposed yet another translation of Kconfig into propositional logic\,\cite{Fernandez-Amoros.ea:2019:kconfig} which, however, omits its three-state logic  and, therefore, a large part of the semantics.

A translation into an SMT representation was created by Xiong et al.\,\cite{rangefix}, who presented RangeFix---a technique to generate configuration conflict fixes specifically for non-propositional configuration spaces. For non-propositional features, it provides ranges to which the value needs to be changed by the user to resolve a conflict. Our fix generation is conceptually based on RangeFix, but through simplifications since we do not need all of its computation steps.

\looseness=-1
A translation not originating from researchers exist as well. As part of a Google Summer of Code project, Vegard Nossum contributed Satconfig\,\cite{satconfig}, which comes with a translation implemented in C and allows reasoning via the SAT solver PicoSAT\,\cite{picosat_perf}, for instance, completing a configuration based on an initial, partial configuration. We investigated the translation of Satconfig as well\,\cite{thesis_daniel}, but found shortcomings in the handling of tristate features, leading to incorrect fixes; there was also limited documentation for the translation and implementation. Despite the limitations, Satconfig was fast, indicating that a C-based solution can be scaled to the size of the kernel's varibility model.
It also showed the advantages of running a SAT solver directly in the configurator tool, as well as the feasibility of implementing the translation into propositional logics in C. Its design inspired our data structures.

With the exception of one tool\,\cite{Fernandez-Amoros.ea:2019:kconfig}, all produce propositional formulas in conjunctive normal form (CNF), a prerequisite for SAT solvers, and typically apply a Tseitin transformation\,\cite{tseytin}, which introduces auxiliary variables to avoid formula explosion.

\looseness=-1
A systematic comparison by ElSharkawy et al.\,\cite{elsharkawy.ea:2015:survey} of LVAT~\cite{lvat}, Undertaker~\cite{undertaker}, and KconfigReader~\cite{kconfigreader} showed that all three tools had shortcomings in their sound encoding and abstraction of the Kconfig semantics. Notably, they found that KconfigReader could correctly handle the majority of the current Kconfig semantics, which steered our decision to re-implement KconfigReader's translation in C with some deviations (explained shortly).

\summary{Kconfig Semantic Abstraction}{%
Over the last decade, multiple researchers and one practitioner implemented propositional abstractions for Kconfig---to provide the basis for SAT-based reasoning, system analysis techniques, and guiding users configuring the kernel. None of these abstractions was fully sound and complete, given the complexity of the Kconfig language. %
}

\parhead{The Kconfig-SAT Initiative}
\noindent
\looseness=-1
In October 2015, the Kernel developer Luis R. Rodriguez contacted some researchers including us who have worked on SAT-based configuration support for the Linux kernel and Linux package management. After discussing configuration issues related to Kconfig, and after being made aware of our research, he launched the Kconfig-SAT initiative, among others, with a project wiki page\,\cite{kconfig-sat-wiki} and a mailing list\,\cite{kconfig-sat-group}. Kconfig-SAT is now also described in the Kconfig documentation\,\cite{kconfig}.

Among the kernel community, the awareness for needing intelligent configuration support rose, despite some skepticism about SAT solvers in general by Linus Torvalds: \textit{``The SAT solver will only hurt, because it will bring in all those irrelevant people who are interested in SAT solving, not in making things easy for users.''}\footnote{\url{https://lists.linuxfoundation.org/pipermail/ksummit-discuss/2017-June/004499.html}}
Nevertheless, our interaction via the mailing list was insightful, as it helped us understanding the kernel community, as well as it provided requirements for a solution.
For instance, in addition to providing sound fixes, a solution should be fast for user acceptance\,\cite{thesis_daniel}.

\summary{Interacting with the Linux Kernel Community}{%
For researchers, it is not clear how the Linux community works and how to effectively interact with it. While we were approached by a kernel developer, for other disseminations, we encourage the community to provide guidelines and requirements for solutions to eventually be integrated, as well as to actively approach researchers for mutual benefit.
}

\section{ConfigFix}
\noindent
We first introduce the notion of configuration conflict, then give a practical introduction into our tool \toolname, followed by discussing its architecture and major design decisions.

\begin{figure}[t]
\centering
	\includegraphics[width=\columnwidth]{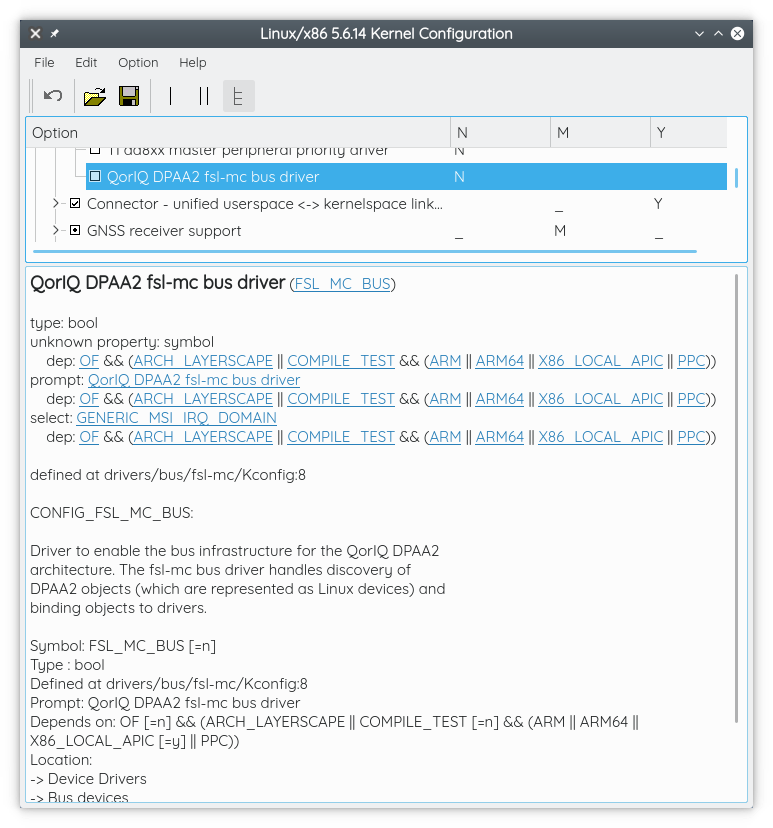}
	\vspace{-.8cm}
	\caption{A configuration conflict in \xconfig} %
	\vspace{-.5cm}
	\label{fig:conflict}
\end{figure}

\subsection{Configuration Conflicts}
\noindent
\looseness=-1
A conflict arises when changing the value of one or multiple features violates a constraint. To resolve conflicts, users need to (transitively) follow the dependencies, which is laborious and error-prone. A survey\,\cite{hubaux_study}
found that users are commonly challenged with conflict resolution in the kernel configurator, with 20\,\% of the survey respondents needing roughly \textquote{a few dozen minutes} to resolve a conflict. The feature descriptions only provide incomplete and sometimes hard to understand (or even incorrect) advice, leading to users blindly choosing default or recommended values without grasping the consequences. Furthermore, default values sometimes contradict with the advice, for instance, when the description recommends enabling a feature, but the default value is ``no'' (disabled). As such, resolving conflicts can be particularly challenging and frustrating for inexperienced users, who lack the experience and resort to guessing.

\looseness=-1
Let us illustrate configuration conflicts with a feature that has particularly complex constraints. \Figref{fig:conflict} shows the feature \textquote{QorIQ DPAA2 fsl-mc bus driver} (a bus infrastructure driver) %
in xconfig. Currently, it cannot be enabled (checkbox not selectable, indicated by missing underscores in columns ``M'' and ``Y'') due to the unmet constraint (current values of those features shown at the bottom of \figref{fig:conflict}):

\begin{smaller}
\begin{verbatim}
     depends on OF && (ARCH_LAYERSCAPE ||
                (COMPILE_TEST && (ARM || ARM64 ||
                X86_LOCAL_APIC || PPC)))
\end{verbatim}
\end{smaller}

\looseness=-1
\Xconfig does not even show which parts of the constraint are unmet. %
The user needs to manually resolve the conflict by transitively looking at the features' dependencies, taking the full richness of its constraint language into account, which might mean needing to enable and disable a set of many features at the same time. Doing so, however, might have ripple effects by triggering the imperative choice propagation (\textsf{select} statement) in \xconfig, which might in turn invalidate the user's resolution. The user might not even realize what other features the imperative choice propagation is enabling or disabling.
Depending on constraints, the respective feature might not even be visible to the users in the configurator, further challenging the configuration process.
The goal of \toolname is to facilitate this conflict resolution process by automatically finding the needed feature values to reach a desired configuration.

\begin{figure}[t]
\centering
\includegraphics[width=.696\columnwidth]{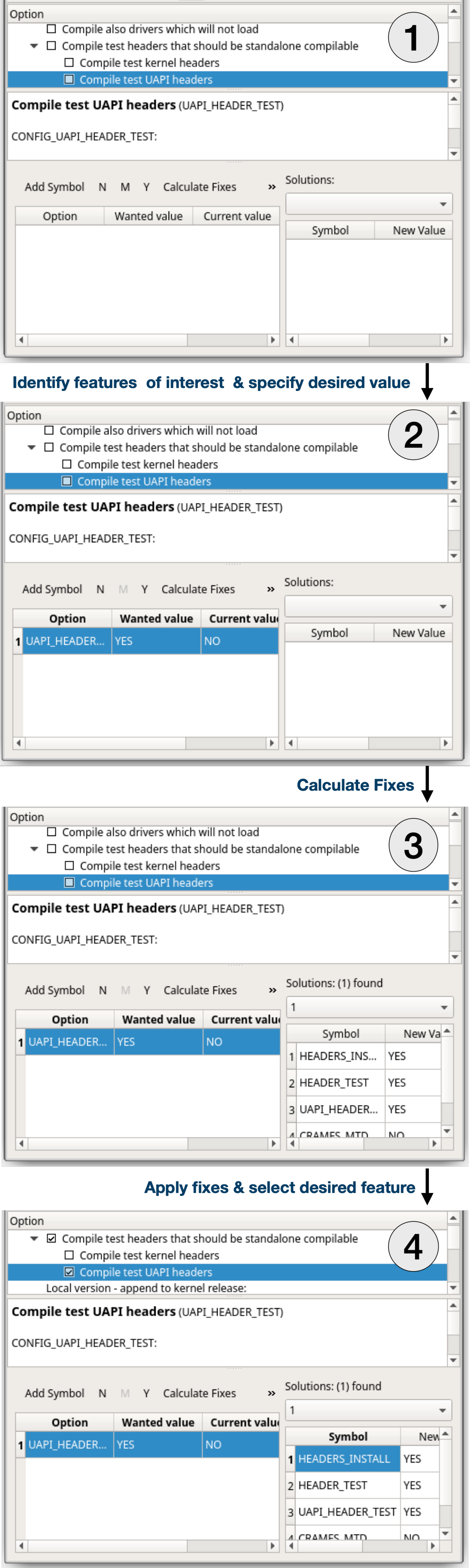}
\vspace{-.2cm}
\caption{User's workflow using \toolname inside \xconfig}
\label{fig:user-workflow3}
\vspace{-.6cm}
\end{figure}

\subsection{\toolname Overview}
\noindent
\parhead{Workflow}
\Figref{fig:user-workflow3} shows the overall workflow of \toolname from the end user's perspective. \circlenum{1} shows\ the view the user sees once they launch \xconfig and have chosen ``show all options'' from the Options menu. This option shows features that would have normally been hidden, because they have unmet dependencies.
The user can then identify the features they wish to enable, and click on ``Add Symbol'' which will add the feature to the bottom left pane, as shown in \circlenum{2}. At that point, the user can change the value of the feature to ``Y'', ``N'' or ``M''---since the feature can be compiled as a loadable kernel module (``M''=module), otherwise only compiled directly into the kernel binary (``Y''=yes) or not at all (``N''=no).
Once the user has added all the features they would like to change the values for, they can click on ``Calculate Fixes,'' which will call \toolname's internal conflict resolution algorithm with the list of features specified in the bottom left pane.
The returned solutions will be displayed in the bottom right pane as shown in \circlenum{3}.
Each solution is a set of feature values that need to be made in order to set the wanted value(s) for the feature(s) indicated in \circlenum{2}.
As there can be multiple ways to satisfy the user constraints, each specific solution
can be viewed by selecting it from the Solutions combobox.
In later versions, we might add feature drag and drop.
The user can then apply any of the solutions, which would allow them to change the values of their desired feature(s).

\parhead{DIMACS Export}
Given the demand for accurate translations of variability models expressed in the Kconfig language, \toolname offers an export into a DIMACS file. DIMACS is a common format accepted by many SAT solvers.\footnote{\url{http://www.satcompetition.org/2009/format-benchmarks2009.html}} The export is launched via \textsf{make cfoutconfig}.

\subsection{Solution Overview}

\begin{figure}[t!]
\centering
	\includegraphics[width=.7\columnwidth]{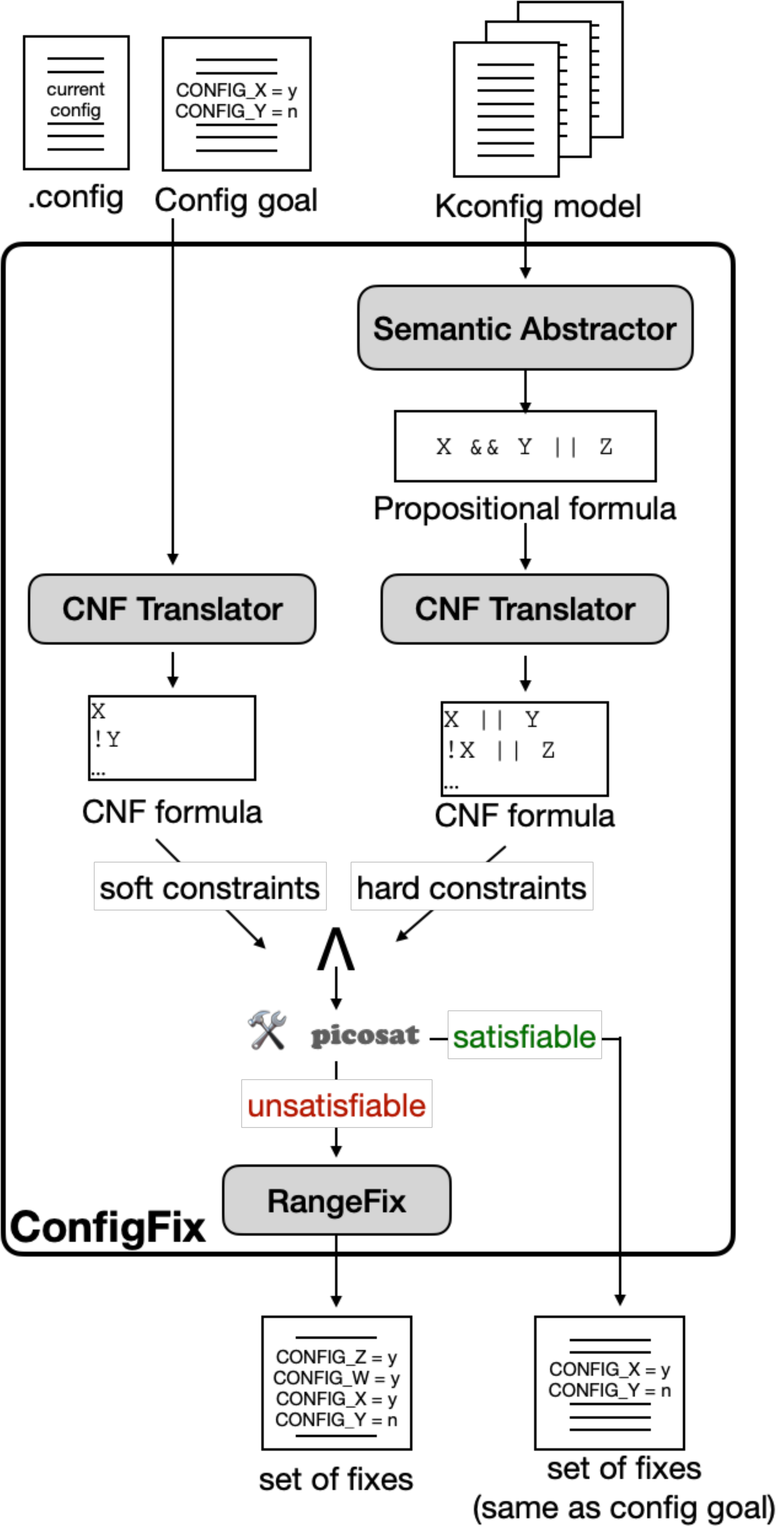}
	\vspace{-.3cm}
	\caption{Overview of ConfigFix's components}
	\label{fig:configfix}
\vspace{-.4cm}
\end{figure}

\noindent
\looseness=-1
\Figref{fig:configfix} illustrates \toolname's fix calculation.
First, \toolname translates the Kconfig model into a propositional formula according to the Kconfig semantics we reverse-engineered and re-implemented (and discuss shortly in \secref{sec:translation}).
It then translates this formula into conjunctive normal form (CNF) to be able to later process it with a SAT solver.
The formula encoding the Kconfig variability model represents the \textit{hard constraints} that must always be satisfied.
In other words, \toolname cannot violate any of these constraints during its fix generation.
As input, \toolname also takes in the current configuration (a list of features and their corresponding values) and the user's configuration goal (a list of features to change and their desired values).
\toolname considers the current configuration as \textit{soft constraints}, since some of the current feature values in the configuration will need to change.
As the next step, \toolname creates a single formula that is a conjunction of all hard and soft constraints and then queries the SAT solver for satisfiability.
If it is satisfiable, then there is nothing to find fixes for, and the desired values can be applied. Otherwise, \toolname triggers our C-based RangeFix implementation to calculate fixes.

\subsection{Fix Generation}\label{sec:rangefix}
\noindent
Given a conflict%
, we want to find a fix that requires a minimal number of changes to the configuration. In the literature, various conflict-resolution algorithms exist\,\cite{hubaux.ea:2012:unifying}, but many are not applicable here. They either produce only one fix, a long list of fixes (challenging users to identify/apply the most suitable one), or they only offer limited support for non-Boolean constraints. We selected \textit{RangeFix}\,\cite{rangefix}, which was designed to mitigate these shortcomings. Its fixes offer a range of values for certain features, and it supports non-Boolean features and constraints.

Rangefix's fixes adhere to three main properties:
\begin{itemize}
	\item \textit{correctness}: any configuration resulting from a fix must be correct, i.e. satisfy the violated constraints;
	\item \textit{maximality of ranges}: when ranges overlap, the fix shall contain a maximum range;
	\item \textit{minimality of changes}: the number of features that need to be changed should be reasonably minimal (realized using heuristics we defined) to avoid unnecessarily breaking features values set by the user before.
\end{itemize}

RangeFix generates fixes in three stages. Its input is an abstraction of a variability model (e.g., CNF formula), where features are represented by variables.
In the first stage, all minimal sets of variables, which have to be changed, are generated. These sets are called diagnoses. In stages 2 and 3, the new values for each variable in a diagnosis are calculated.

We now explain the algorithm using an example.
Let us define the tuple $(V,e,c)$ as a \textit{constraint violation}, where $V$ is a set of variables, $e$ a configuration with a value defined for each variable and $c$ a set of constraints which is violated. The goal is to find a new configuration $e'$, such that $c$ is satisfied. We define our set of variables $V$ as:
$$\{m: Boolean, a: Integer, b: Integer\}$$
We define a set of constraints as:
$$(m \rightarrow a > 10)\ \land (\neg m \rightarrow b > 10)\ \land (a < b)$$
Finally, a configuration $e$ with values for the three variables is required:
$\{m:= True, a:= 6 , b:= 5\}$. This configuration violates first and the third constraint.

RangeFix generates the diagnoses during the first stage by using the constraint solver's ability to find unsatisfiable cores. During each iteration in the first step, a constraint from an unsatisfiable core is removed and the diagnoses are extended until no more unsatisfiable core is found. Applying this to our example yields the following two diagnoses: $\{m, b\}$ and $\{a, b\}$. At this point, we know that we need to change either $\{m, b\}$ or $\{a, b\}$ to resolve the conflict.

The stages 2 and 3 are subsequently performed together for each diagnosis. First, all unchanged variables are replaced by their current values during stage 2. Finally, in stage 3, the violated constraints are minimized via heuristic rules we defined and split into minimal clauses to generate the fixes. This leads to the following two fixes for the conflict:
\begin{itemize}
	\item $[m:= False, b: b > 10]$
	\item $[(a,b): a > 10 \land a < b]$
\end{itemize}

\looseness=-1
If any of these two fixes is applied, all previously violated constraints will be satisfied again. If the first fix is chosen, then $m$ needs to be set to $False$ and $b$ simply to any value larger than 10. If the second fix is chosen, then $a$ needs to be set to a value larger than 10, and $b$ must simply be larger than $a$. This illustrates the advantages of RangeFix compared to existing conflict-resolution algorithms. While there are infinitely many possible solutions, the user is presented with only the minimal set. Other approaches might have presented only one of the two solutions or a very long list of possible solutions, since there are infinitely many possible combinations that satisfy $a$ being smaller than $b$ in the second fix.
\looseness=-1
As such, \toolname returns a maximum of three fixes for each conflict even if more fixes exist. In most cases, presenting three fixes for a user is sufficient to choose a suitable fix and generating more fixes can take significantly more time.

\subsection{Example Fix}\label{sec:examplefix}
\looseness=-1
\noindent
We now demonstrate how \toolname finds different fixes for a conflict, and how it deals with the intricacies of Kconfig's semantics, especially its various language constructs to define constraints (cf. \secref{sec:kconfigconcepts}). %
Given a default configuration, assume a user wants to enable the feature \texttt{MEDIA\_TUNER\_SIMPLE}, which provides support for various media tuners. The feature has dependencies, and its parent feature %
depends on other features as it has visibility constraints. In the default configuration, the parent is hidden and the dependencies are not met, so it is not configurable by the user.

For this conflict, eight possible fixes exist and,
each fix changes the values of five features, although some values change implicitly through Kconfig's imperative choice propagation (\texttt{select} statements). Generally, the feature can be enabled through two different means: The parent feature can be made visible and then the user can explicitly set a value for \texttt{MEDIA\_TUNER\_SIMPLE} after its dependencies have been met. The other possibility is to enable the module through a \texttt{select} statement while keeping the module invisible.

In the first case, the following fix first makes the feature visible, and then the user can set an explicit value:

\begin{smaller}
\begin{itemize}
	\item \texttt{MEDIA\_SUPPORT => yes}
	\item \texttt{MEDIA\_DIGITAL\_TV\_SUPPORT => yes}
	\item \texttt{MEDIA\_SUBDRV\_AUTOSELECT => no}
	\item \texttt{MEDIA\_TUNER\_SIMPLE => yes}
\end{itemize}
\end{smaller}

In the latter case, the following fix ensures that the module is selected and, therefore, enabled:

\begin{smaller}
\begin{itemize}
	\item \texttt{MEDIA\_SUPPORT => yes}
	\item \texttt{MEDIA\_ANALOG\_TV\_SUPPORT => yes}
\end{itemize}
\end{smaller}

The feature is invisible, though, and it might not be obvious to the user why the feature has been enabled.
In the end, all eight fixes calculated are able to enable \texttt{MEDIA\_TUNER\_SIMPLE} in one way or another.

\section{Experiences and Challenges}
\label{sec:challenges}
\noindent
\looseness=-1
We now discuss experiences and challenges faced when realizing \toolname. %
Overall, our first attempt\,\cite{thesis_daniel} was much closer to the RangeFix implementation. We experimented with the C-based translation by Vegard Nossum (cf. \secref{sec:translations}) and an existing Scala-based RangeFix implementation prototypically integrated into \xconfig, but which only covered the first of the three stages of RangeFix. We surveyed kernel developers and Kconfig-SAT members about the user interface we implemented, and provided screencasts illustrating the solution\,\cite{thesis_daniel}.
The results contributed to the present attempt, where we focused on a purely C-based translation and fix generation, the C-based SAT solver Picosat\,\cite{picosat}, and an improved integration into \xconfig.

\subsection{Semantic Abstractor}\label{sec:translation}  %
\noindent
\looseness=-1
The largest challenge was obtaining a sound and stable logical abstraction of Kconfig. %
To obtain requirements, we interacted with the community, specifically via the kconfig-sat mailing list. We recognized a strong preference for SAT solvers, as opposed to more expressive solvers, such as SMT. Even though, the latter could support a larger part of the semantics, it was pointed out that integrating a SAT solver could also help at other places in the kernel, especially CPU scheduling support. Furthermore, SMT solvers are typically slower, and not many come with a GPL-compatible license, as required for integration into the kernel's codebase.
For practical matters, we learned it should be possible to use or compile the SAT solver with the tools needed to compile the Linux kernel, such as \textit{gcc}. This requirement excluded some modern and very fast solvers, including CaDiCaL\,\cite{cadical}.

In general, our options for realizing a propositional semantic abstraction for a SAT solver were to (i) develop a new translation from scratch, to (ii) build on \textit{Satconfig} (cf. \secref{sec:translations}) or to (iii) investigate other existing alternatives. Strategy (i) has the disadvantage of lacking a reference, when the translation is checked for correctness. Since correctness was essential, we disregarded this strategy. As explained above, Satconfig showed deficiencies in handling \texttt{tristate} features and in code documentation. Based on others' systematic comparison of translations\,\cite{elsharkawy.ea:2015:survey} (also cf. \secref{sec:translations}), we chose to re-implement and extend the Scala-based \textit{KconfigReader}\,\cite{kconfigreader}.

\looseness=-1
Our overall strategy was to inspect the translation in Kconfig\-Reader, to re-implement it freely in C, being inspired by data structures from Satconfig, and to test the translation incrementally with smaller, hand-crafted models, thereby also debugging and fixing our implementation as well as remaining deficiencies of KconfigReader.
Finally, we tested with example conflicts in the full kernel model (one of these conflicts we described in \secref{sec:examplefix}).
KconfigReader works accurately for \texttt{bool} and \texttt{tristate} features, as well as for many non-Boolean properties, but with some remaining smaller limitations.

\looseness=-1
Adapting and re-implementing KconfigReader in C allowed the direct integration into the kernel configurator in the kernel source tree. It also allowed us to use the configurator's parser to parse the variability model. In fact, parsing Kconfig is another general challenge acknowledged by the researchers who produced translation. So, relying on the maintained parser provides robustness. We traversed the internal, AST-based representation and stored intermediate results in our own C data structures. A specific challenge was to implement scalable representations of the propositional formula in C, in a way that it can be easily traversed and transformed into conjunctive normal form by applying the typical logical laws and a Tseitin transformation.
As a consequence, some parts of the translation had to be implemented in a completely different way than in KconfigReader.

\summary{Propositional Transformation in C}{%
C was not ideal to model propositional logic formulas. An OO language, especially one with functional-programming constructs, would have likely led to much cleaner code.
}

\looseness=-1
To account for limitations, we needed to deviate from the semantics realized in KconfigReader. One example was the translation of the imperative choice propagation---over 10,000 \texttt{select} statements exist in the whole variability model. So, how to model this type of constraint has substantial impact on the performance.
KConfigReader models the \texttt{select}-constraints under the constraints for the selected feature and not the selecting feature. An advantage of this behavior is the number of constraints, since there will only be one constraint for each selected feature independently by how many other features it is selected. The disadvantage is that the constraints can potentially become very large formulas, when a feature is selected by many other features and depending on how many constraints these features have.

\summary{Challenges of Achieving a Sound Semantic Abstraction}{%
As known for over a decade, the Kconfig syntax and semantics are intricate. Especially, the semantics are more expressive than logical representations of off-the-shelf reasoners, including SMT and of course SAT. In addition, the continuous evolution of Kconfig, lack in documentation, and non-obvious semantic abstractions, challenged providing a sound abstraction. The kernel configurator's code is also not well documented, only some of its data structures, no functions. Furthermore, a naive CNF translation explodes without Tseitin transformation and incorporating domain/expert knowledge for effectively splitting clauses.
}

\subsection{Choosing a SAT Solver}
\noindent
Choosing PicoSAT\,\cite{picosat} was inspired by Satconfig (cf. \secref{sec:translations}), which showed that PicoSAT can be easily integrated into the kernel. More importantly, it:	(i) is written in C and can be compiled with \texttt{gcc}; (ii) has a C-API, so can be called called directly within \xconfig without needing external calls; (iii) has a Linux-compatible license (MIT); (iv) can identify and return unsatisfiable cores; and (v) is reasonably fast\,\cite{picosat_perf}.
This made PicoSAT our best candidate. However, a downside was that the solver is outdated and not actively maintained  anymore, despite still being use in industry.

\summary{Choice of SAT Solver}{%
While faster SAT solvers exist, the need to use a pure C-based SAT solver with a C API and Linux-kernel-compatible license, restricted our options to PicoSAT. While it is not actively maintained anymore, it has a well-structured implementation and turned out to be fast enough for interactive fix generation in \toolname.
}

\subsection{GUI Integration}
\looseness=-1
\noindent
We extended the graphical configurator \xconfig to provide an intuitive interface for entering the desired feature values, for observing the proposed fixes, and for applying the desired ones.
We decided on having the conflict resolution integrated within the same window to follow conflict resolution interfaces found in other configurators, specifically that of \ecos\,\cite{veer.ea:2001:cdl,hubaux_study,berger2010variability}.
We added a new pane at the bottom of \xconfig to collect the features desired by the user. The view is divided into two parts: (i) the collected feature list and (ii) the solutions obtained from \toolname. We initially planned to support drag and drop, left it for later in favor of obtaining a working solution for now.

After realizing the initial UI, we started realizing the interfaces between the UI, our RangeFix implementation%
, and the SAT solver PicoSAT via its C API.
\xconfig is written in C++ with the Qt toolkit, while we implemented \toolname in pure C and call it within \xconfig.
The Glib library %
is used in \toolname and \xconfig for its variety of available data structures resulting. %

The challenge with implementing and integrating the user interface was in understanding the libraries Glib and Qt, specifically Qt slots and signals, and the type-checked event signalling mechanism in Qt. Furthermore, we needed to figure out how to interface C++ with C code (name unmangling), and how to pass data around between user interface elements (parent-child and sibling in the interface hierarchy).

\subsection{Scalability and Performance Improvements}
\noindent
\looseness=-1
The main challenge was translating the huge formulas of the full kernel variability model with over 15,000 features into CNF. Notably, the encoding of certain Kconfig aspects had a substantial impact on the resulting CNF and SAT solver performance.

\looseness=-1
To improve the performance of the SAT solver, we changed the encoding of the \texttt{select} statement by splitting up the various statements. Instead of a single constraint for a selected feature, each \texttt{select} statement now creates constraints on its own. But, since this also had an affect on we encoded the dependencies of a feature, introducing a new variable was needed, which indicates whether a feature has been selected. This significantly simplified the constraint encoding.

\looseness=-1
While we have increased the number of constraints in total by several thousand constraints, we were able to reduce the number of CNF clauses and auxiliary variables significantly As a consequence, a single run of PicoSAT became more than 65\,\% faster than it was before. A disadvantage of this decision is, that we lost the ability to syntactically check our constraints for equivalence against the constraints produced by KConfigReader. Still, the gain in performance justified this decision, and we conceived an alternative evaluation, explained shortly.

\summary{Achieved Performance}{%
We achieved a final translation time of the entire Linux kernel variability model into a CNF formula of around 1.5 seconds on an Intel i7 laptop. The initial run of PicoSAT to check for satisfiability takes about 2.5 to 3 seconds. Finally, finding fixes for a conflict can be achieved in as little as 1 second in some cases, although the number depends heavily on the conflict and the number of enabled features. The most impact on improving the performance came from incorporating our domain/expert knowledge into the translation, including effective formula splitting and using a Tseitin transformation.
}

\section{Evaluation}
\label{sec:evaluation}
\noindent
\looseness=-1
We now briefly discuss our evaluation of \toolname. %

\begin{figure*}[t]
\centering
	\includegraphics[width=\linewidth]{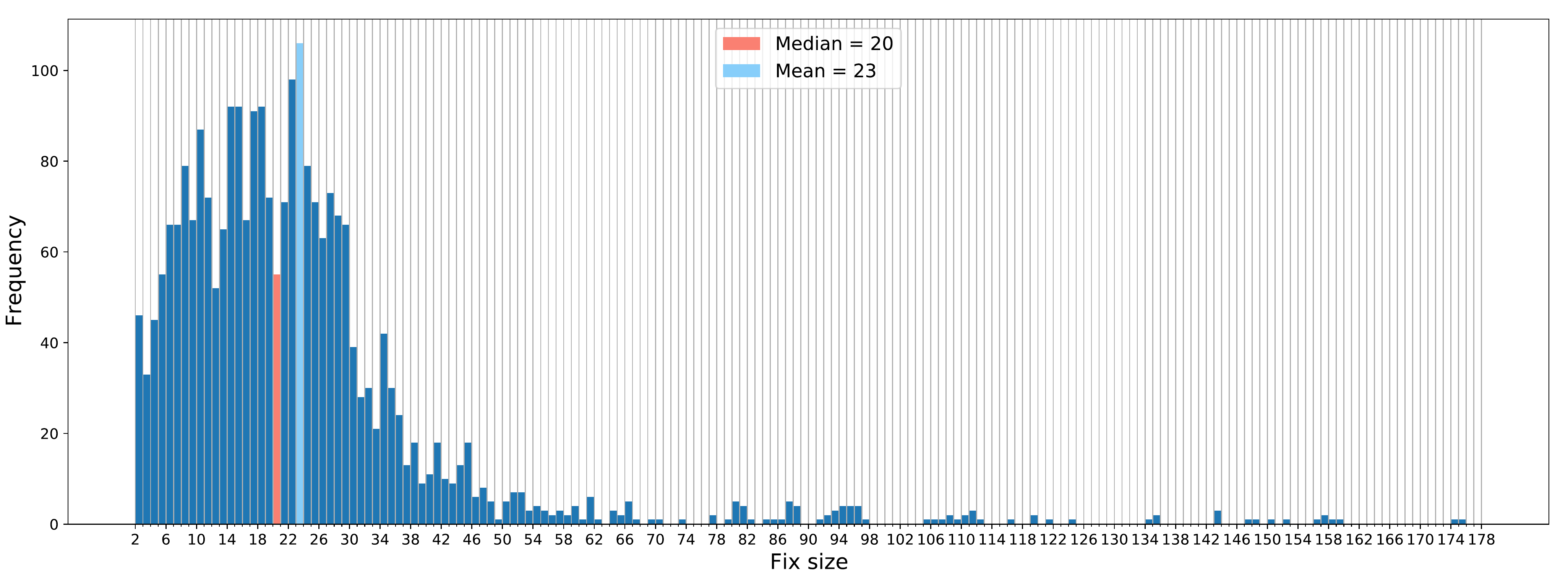}
	\vspace{-.5cm}
	\caption{Distribution of fix sizes (number of features that need to be changed to resolve the conflict)}
	\label{fig:histogram}
\end{figure*}

\subsection{Conflict and Fix Generation}
\noindent
\looseness=-1
With over 15,000 features, the configuration space for the Linux kernel is huge and, therefore, crafting a small number of examples to be evaluated is not sufficient. Instead, a more systematic approach is needed. We made use of the kernel's ability to randomly sample a configuration and then we created random conflicts (that are definitely resolvable, see below) for each sampled configuration.

To obtain sufficiently diverse configurations, we used the kernel's \texttt{randconfig} tool to generate configuration samples for three of the more popular available architectures in the kernel: \texttt{x86\_64}, \texttt{arm64}, and \texttt{openrisc}.
Randconfig allows to skew the probabilities for features of type \texttt{bool} and \texttt{tristate} to be enabled.
Our probabilities ranged from 10\,\% to 90\,\% for a feature to be set to \texttt{no}. 
This setting affects the number of enabled features and, consequently, the fix sizes (number of features to be changed), since a low number of enabled features tends to lead to larger fixes regardless of the conflict size. This way, we can evaluate the performance of \toolname for varying fix sizes.

\looseness=-1
For each random configuration we then introduce conflicts by randomly choosing target features that: (i) have a prompt (i.e., are not completely invisible, like derived features\,\cite{berger2013study}), (ii) are of type \texttt{bool} or \texttt{tristate}, (iii) are not a choice group, and (iv) have at least one other possible value different from its current value that cannot be selected at the moment. In the case of \texttt{tristate} options with two values that cannot be selected at the moment, the target value is randomly chosen from these two values.

\looseness=-1
While finding fixes, \toolname may not return results for two reasons. The first is that a fix may not exist. 
Some features depend on other architecture-specific options; therefore, they can only be configured for certain architectures. 
Thus, 
including such architecture-dependent features into conflicts to be resolved 
on a different architecture by definition has no fix.
In this case, \toolname not returning a fix is the expected behavior.
On the other hand, \toolname may also not return fixes due to bugs in our implementation or design. In order to objectively evaluate \toolname, we, therefore, want to ensure that our chosen target features can be configured (i.e., the conflict can be resolved) for the used architecture to rule out the first possibility. 
Therefore, we generate a base configuration for each architecture using \texttt {randconfig} with the probability of 100\% for a feature to be enabled. Such base configuration will have as many enabled features as possible. 
Before including a feature in a conflict, we compare its target value with the base configuration to see whether this value can indeed be obtained for that architecture.

\looseness=-1
For each configuration sample, we restrict the selection of conflicting features to those that can receive their target values on the chosen architecture, as witnessed by the base configuration. We create such resolvable conflicts containing between 1--10 features and let \toolname generate fixes for each conflict.

\subsection{Analysis}
\noindent
When applying each produced fix for a given conflict, we analyze whether (i) the fix resolves the conflict, i.e. whether our randomly chosen features obtain their target values and whether (ii) every feature in a fix obtains its target value if the fix resolves the conflict after applying the fix. This results in three possible outcomes for each fix:

\begin{figure}[t]
\centering
	\includegraphics[width=\columnwidth]{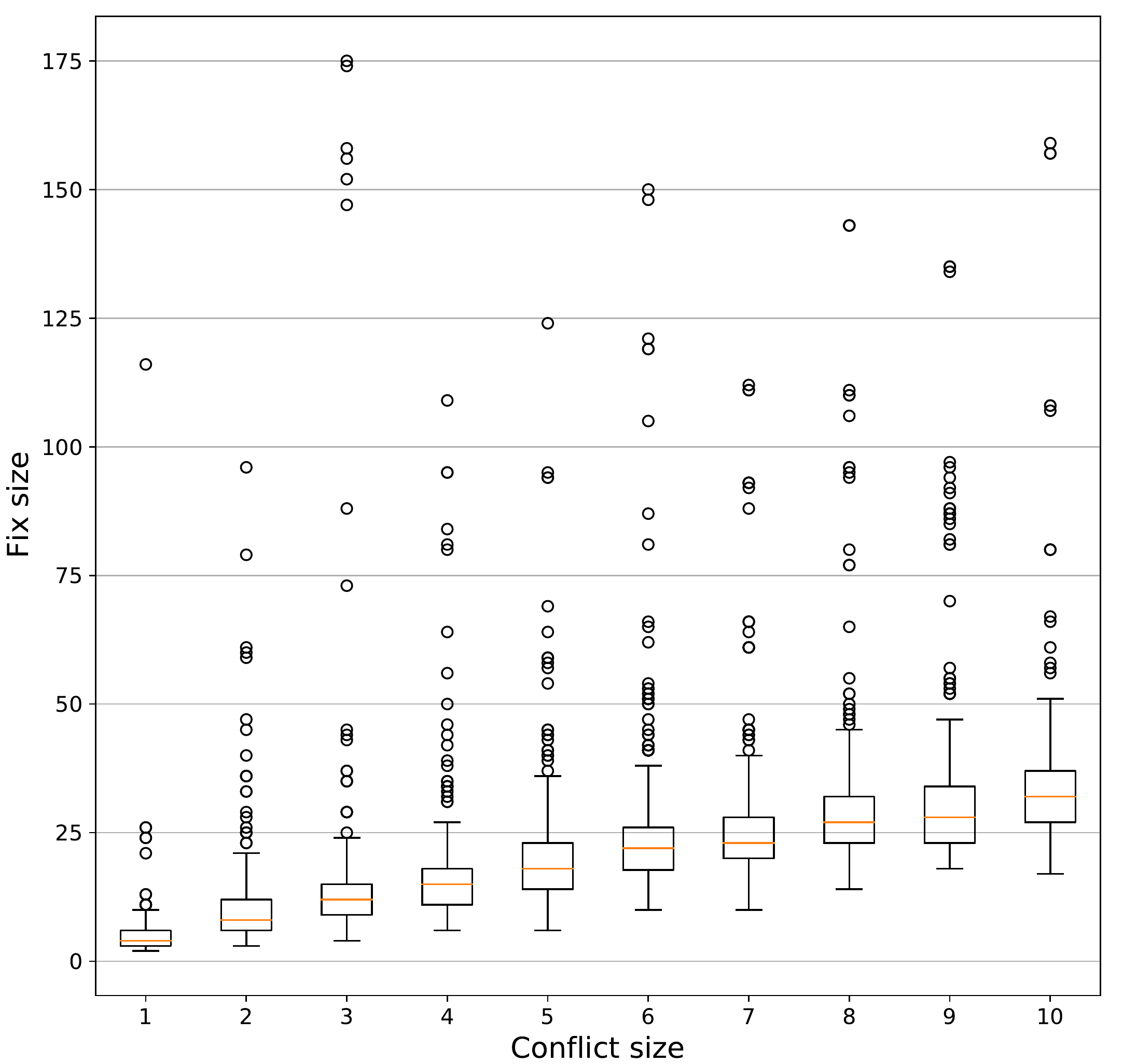}
	\vspace{-.5cm}
	\caption{Distribution of fix sizes (number of features that need to be changed) by conflict size (number of unchangeable features that a user wants to change)}
	\label{fig:boxplot}
\end{figure}

\begin{itemize}
	\item \textbf{Applicable and Resolves Conflict:} The fix is fully applicable (i.e., we can apply all the specified values in the fix) and resolves the conflict. This is the optimal outcome.
	\item \textbf{Not Applicable, but Resolves Conflict:} The fix is not fully applicable (i.e., some of the values specified in the fix cannot be applied), but it resolves the conflict nonetheless. While the fix is not optimal, since it contains invalid or redundant feature values, it is an acceptable outcome, since the conflict is resolved. Specifically, the features to change, which are part of the fix, can be changed.
	\item \textbf{Does Not Resolve Conflict:} The proposed fix does not resolve the conflict, since the target features cannot be changed by applying the conflict.
\end{itemize}

\subsection{Results}
\noindent
We generated a total of 27 random sample configurations with a varying number of enabled features for the three target architectures (\texttt{x86\_64}, \texttt{arm64}, and \texttt{openrisc}). For each configuration, we created 50 conflicts containing between 1--10 features to change. This resulted in 1350 conflicts for \toolname to solve.
We summarize the results in \tabref{tab:stats}.%

Out of the 1,350 conflicts, \toolname returned at least one fix for 1,055 (78.2\,\%) conflicts, of which 723 (53.6\%) were resolved. The remaining 295 conflicts did not receive any fixes, mostly due to a limit on the \toolname running time, which we used to make testing feasible.
For the 1,055 conflicts that received at least one fix, a total of 2,482 fixes were returned (recall our intentional limitation to a maximum of three fixes per conflict, as discussed in \secref{sec:rangefix}).

As seen in \figref{fig:histogram}, the obtained fixes comprise 2--175 features that need to be changed, although the typical range is much smaller. The distribution is left-skewed, with an average fix size of 23 features, and the majority of the fixes lying around the median of 20 features. \Figref{fig:boxplot} shows that outliers are common for all conflict sizes, but in general, fix sizes demonstrate a seemingly linear dependence on the conflict size. This indicates that \toolname can be used for resolving conflicts of varying sizes without the risk of fix size explosion. 

We also investigated the outcome of each fix, summarized in \tabref{tab:stats}. The optimal outcome (\textbf{Applicable and Resolves Conflict}) was achieved for 1,317 of these fixes (53\,\%), while 292 fixes (12\,\%) still resolved the conflict despite not being fully applicable (\textbf{Not Applicable, but Resolves Conflict}). Finally, 868 (35\,\%) returned fixes that did not resolve the conflict (\textbf{Does Not Resolve Conflict}). So, in summary, 65\,\% of the fixes returned by \toolname resolved the conflict.

\summary{Evaluation Summary}{
\noindent
We evaluated \toolname with 1,350 conflicts of different sizes that were randomly generated for 27 configuration samples on three hardware architectures. 65\,\% of the fixes generated by \toolname resolved the conflict.
}

\looseness=-1
While the ideal outcome would have been to resolve all conflicts, we believe that this percentage is still acceptable given that the semantic abstraction is sound, but cannot be complete. While perhaps an encoding into SMT could yield a slightly better result, recall that there was a strong preference amongst the Linux community for SAT solving using a C-based solver.

\begin{threeparttable}[b] 
  \caption{\toolname evaluation results for random configurations and conflicts}%
  \label{tab:stats}
	\begin{smaller}
  \begin{tabularx}{\linewidth}{ll}
    \toprule
    \textsf{metric} & \textsf{value}\\
    \midrule
    number of sampled configurations 							& 27\\
	  conflict sizes 			  							& 1--10\\
    generated conflicts   								& 1,350\tnote{1} (100.0\,\%)\\ %
		\midrule
	conflicts with at least one generated fix  								& 1,055 (78.2\,\%)\\
	number of resolved conflicts			& 723 (53.6\,\%) \\  %
	total number of generated fixes										& 2,482 (100.0\,\%)\\
	\textbf{fixes that resolve the conflict} & \textbf{1,609 (65.0\,\%)} \\
	\hspace{.4cm} fully applicable and resolve the conflict	& 1,317 (53.0\,\%)\\
	\hspace{.4cm} not fully applicable, but resolve the conflict & 292 (12.0\,\%)\\
	\hspace{.4cm} do not resolve the conflict							& 868 (35.0\,\%)\\
    \bottomrule
  \end{tabularx}
\begin{tablenotes}[para]
\item[1] For each configuration sample, five conflicts of each size.
\end{tablenotes}
\end{smaller}
\end{threeparttable}

\section{Conclusion}
\label{sec:conclusion}
\noindent
\looseness=-1
We reported our experience of leveraging results from 13 years of academic research. We created the tool \toolname, which we integrated in the Linux kernel configurator by adhering to all requirements coming from the Linux community in the context of the Kconfig-SAT initiative\,\cite{kconfig-sat-wiki}. \toolname realizes a stable and sound transformation of the configurator's underlying variability-modeling language Kconfig into a propositional abstraction, as well as it provides a tested configuration-conflict resolution technique that can guide users achieving their desired Linux kernel configuration. We believe that our tool\,\cite{configfix} helps the Linux kernel community not only supporting the configuration process, but also conduct further analyses or support based on the stable translation. Likewise, we invite researchers evaluating their own and novel techniques upon the translation, as well as to improve the fixes, for instance, providing optimization support based on quality attributes.

\bibliographystyle{IEEEtranS}
\bibliography{doc}

\end{document}